\documentclass[letterpaper]{iopart}

\usepackage{graphics,graphicx,epsfig,bbm}
\usepackage{amsthm,amsfonts,amssymb,amscd,latexsym,color}

\begin{document}

\def\qed{$\Box$}
\def\setminus{\smallsetminus}
\def\ens#1{\{#1\}}
\def\ie{\textit{i.e.}}
\def\eg{\textit{e.g.}}
\def\tr#1{\textcolor{red}{#1}}
\def\mathvec#1{\mathbf{#1}}
\def\al{\alpha}
\def\ba{\beta} 
\def\rar{\rightarrow}
\def\lar{\leftarrow}
\def\srar{\mathrel{\rar\hskip-1.95ex\rar}}
\def\Rar{\Rightarrow}
\def\Lar{\Leftarrow}
\def\lrar{\longrightarrow}
\def\lRar{\Longrightarrow}
\def\Lrar{\Leftrightarrow}
\def\lRar{\leftrightarrow}
\def\mcl{\mathcal} 
\def\mbb{\mathbb} 
\def\mfr{\mathfrak}
\def\MA#1{\left(\begin{matrix}#1\end{matrix}\right)}
\def\EQ#1{\begin{eqnarray}#1\end{eqnarray}}
\def\ket#1{{|}#1\rangle}
\def\bra#1{\langle#1{|}}
\def\ctwo{{\mbb C}^2}
\def\ztwo{{\mbb Z}_2}
\def\pit{{\begin{matrix}\frac\pi2\end{matrix}}}
\def\pif{{\frac\pi4}}
\def\oqb#1{\ket{\hskip-.1ex+_{#1}}}
\def\oqbn#1{\ket{\hskip-.1ex-_{#1}}}
\newtheorem*{props}{Proposition}\def\PRO{\begin{prop}}\def\ORP{\end{prop}}
\newtheorem*{coros}{Corollary}\def\COR{\begin{coro}}\def\ROC{\end{coro}}
\newtheorem{prop}{Proposition}\def\PRO{\begin{prop}}\def\ORP{\end{prop}}
\newtheorem{coro}{Corollary}\def\COR{\begin{coro}}\def\ROC{\end{coro}}
\newtheorem{theo}{Theorem}\def\TH{\begin{theo}}\def\HT{\end{theo}}
\newtheorem{prob}{Problem}\def\PRB{\begin{prob}}\def\BRP{\end{prob}}
\newtheorem{defi}[prop]{Definition}\def\DE{\begin{defi}}\def\ED{\end{defi}}
\newtheorem{lemme}[prop]{Lemma}\def\LE{\begin{lemme}}\def\EL{\end{lemme}}
\newcommand{\AR}[2][c]{$$\begin{array}[#1]{lllllllllllllll}#2\end{array}$$}
\newcommand{\enc}[1]{\overline{#1}} 
\newcommand{\cptp}[1]{\mathbb{#1}}
\newcommand{\gt}[1]{\mathtt{#1}}  


\def\newQ#1#2{N_{#1}^{#2}}
\def\ctR{\mathop{\wedge}\hskip-.4ex} 
\def\cx#1#2{X_{#1}^{#2}}
\def\cz#1#2{Z_{#1}^{#2}}
\def\cp#1#2#3{Z(#1)_{#2}^{#3}}
\def\czp#1#2{P_{#1}^{#2}}
\def\czpp#1{Q_{#1}}
\def\mLR#1#2#3#4{{}_{#4}[{M}_{#2}^{#1}]^{#3}}
\def\mR#1#2#3{\mLR{#1}{#2}{#3}{}}
\def\mL#1#2#3{\mLR{#1}{#2}{}{#3}}
\def\m#1#2{{M}_{#2}^{#1}}
\def\et#1#2{E_{#1#2}}

\newcommand{\poe}{\Omega}

\newcommand{\eqdef}{\triangleq}

\newcommand{\tfrac}[2]{\case{#1}{#2}}
\newcommand{\eqref}[1]{\eref{#1}}

\title[A direct approach to FT  
in measurement-based QC via teleportation]
{A direct approach to fault-tolerance  
in measurement-based quantum computation via teleportation}

\author{Marcus Silva$^{1,2}$, Vincent Danos$^3$, Elham Kashefi$^{1,4}$, Harold Ollivier$^5$}
\address{
$^1$Institute for Quantum Computing, U. of Waterloo, 200 University Ave. West, Waterloo, ON, N2L 3G1, Canada}
\address{
$^2$Department of Physics and Astronomy, U. of Waterloo, 200 University Ave. West, Waterloo, ON, N2L 3G1, Canada}
\address{
$^3$Universit\'e Denis Diderot \& CNRS, 175 Rue du Chevaleret, 75013 Paris, France}
\address{
$^4$Christ Church College, University of Oxford, OX1 1DP, Oxford, UK}
\address{
$^5$Perimeter Institute for Theoretical Physics, 31 Caroline Street North, Waterloo, ON, N2L 2Y5, Canada}
\ead{\mailto{msilva@iqc.ca}}


\begin{abstract}
We discuss a simple variant of the one-way quantum computing 
model~[R. Raussendorf and H.-J. Briegel, PRL 86, 5188, 2001],
called the Pauli measurement model, where 
measurements are restricted to be along the eigenbases of
the Pauli $X$ and $Y$ operators, while qubits can be initially prepared both
in the $\ket{+_{\pi\over 4}}:={1/\sqrt{2}}(\ket{0}+e^{i{\pi\over 4}}\ket{1})$ state
and the usual $\ket{+}:={1/ \sqrt{2}}(\ket{0}+\ket{1})$ state.
We prove the universality of this quantum computation model, and
establish a standardization  procedure which permits all entanglement
and state preparation to be performed at the beginning of computation. This
leads us to develop a direct approach to fault-tolerance by simple
transformations of the entanglement graph and preparation operations,
while error correction is performed naturally via
syndrome-extracting teleportations.
\end{abstract}

\maketitle


\section{Introduction}

The one-way quantum computation (1WQC) model~\cite{RB01} has been
widely studied since its discovery. One particular issue that has
attracted attention is how to perform fault-tolerant (FT) quantum
computation (QC) in such a model~\cite{AGP05,Rau03,ND05,AL05,RHG05,RH06}.
While FTQC can be performed through
such a model by simulating FT quantum circuits via 1WQC~\cite{ND05,AL05},
FTQC can also be achieve directly through the use of topological
error correction techniques~\cite{RHG05,RH06}.
The focus of this paper is to illustrate another direct approach to FTQC
in measurement based computation, building on insights into
the measurement calculus~\cite{generator04} and generalizations of 1WQC, 
as well as teleportation-based approaches to error correction~\cite{Kni05b,Kni05a}.

We consider a model where measurements in the full
$XY$-plane are traded off for more complex preparations of the
vertices in the entangled resource state. This model, which we call the
{\em Pauli measurement model} (PMM), uses only measurements
along the $X$ and the $Y$ directions, while the entangled resource state
is obtained via initialization of individual qubits into the state
$\ket{+}:=\tfrac{1}{\sqrt{2}}(\ket{0}+\ket{1})$ or 
$\ket{+_{\pi\over 4}}:=\tfrac{1}{\sqrt{2}}(\ket{0}+e^{i{\pi\over 4}}\ket{1})$,
followed by application of the unitary interaction $\ctR{Z}:={\mathrm{diag}}(1,1,1,-1)$
(also known as the controlled-$Z$ gate)
between certain pairs of qubits. We show that the PMM model is fault-tolerant in the 
usual simulation sense~\cite{ND05,AL05}. Moreover, through the use of
encoded or nested graph states~\cite{Dan05a}, and the careful selection
of quantum codes, all necessary operations for computation can be performed
transversally on encoded information, so that the graph state computation
itself is made fault-tolerant if the error rate is low enough.

First, we investigate how to extend the main properties 
of the 1WQC model using these modified preparation states, while still maintaining the 
properties one needs for convenient error correction. We then demonstrate that this 
model naturally provides the resources necessary for fault-tolerant syndrome extraction, 
and illustrate how any PMM computation can be transformed into
a larger one that has a lower effective error rate if the error rate per
operation is below some threshold, achieving fault-tolerance.


\section{One-way quantum computation with phase preparation} 

Consider a slight extenstion to 1WQC where
a measurement pattern, or simply a pattern, is defined by a sequence of quantum operations
over a finite set of qubits $V$, along with two subsets $I\subseteq V$ and
$O\subseteq V$ representing the pattern inputs and outputs
respectively ($I$ and $O$ may intersect).  
The allowed operations are: 
(a) $N_i^\alpha$, preparation of qubit $i$ in the state 
$\ket{+_\alpha}=\tfrac{1}{\sqrt{2}}(\ket{0}+e^{i\alpha}\ket{1})$; 
(b) $E_{ij}$, unitary interaction between qubits $i$, $j$ of the form $\ctR Z$;
(c) $M_i^\alpha$, measurement of qubit $i\not\in O$ in the 
$\ket{\pm_\alpha}=\tfrac{1}{\sqrt{2}}(\ket{0}\pm e^{i\alpha}\ket{1})$ eigenbasis, with outcome $s_i\in\{0,1\}$ corresponding 
to collapse into the state $\ket{+_\alpha}$ or $\ket{-_\alpha}$, respectively;
(d) $X_j$ and $Z_j(\alpha):=e^{-i{\alpha\over 2}Z_j}$, local unitaries on qubit $j$.
In addition, local unitaries and measurement basis may depend on the outcome of measurements of other
qubits, which is denoted in the natural way, e.g. $X_j^{s_k}$ indicating a unitary which acts if $s_k=1$, 
or $M_j^{\alpha-s_k\beta}$ indicating a measurement in a basis which depends on the measurement outcome $s_k$.  

Measurements are considered to be destructive, and we require that no
operations be performed on measured qubits. We also only consider runnable 
patterns where no operations depend on the outcome of measurements that have 
not yet been performed. 
Local unitaries are crucial for the understanding of how universality
and determinism come about (recall that measurement outcomes in
quantum mechanics are, in general, non-deterministic)~\cite{Rau03,generator04}.
Both 1WQC and PMM are particular cases of this more general model: to
obtain the 1WQC model set $\alpha=0$ in clause (a); to obtain the PMM,
set $\alpha=0,\pi/4$ in clause (a), and $\alpha=n\pi/2$ in clauses (c)
and (d). 

Patterns, denoted by gothic letters, e.g.\ ${\mfr A}$ and
${\mfr B}$,  can be combined to create a new pattern via parallel
concatenation ${\mfr A}\|{\mfr B}$, or serial concatenation 
${\mfr A}\circ{\mfr B}$.  Parallel concatenation means the qubits are
relabelled in such a way that all operations in ${\mfr A}$ commute with
all the operations in ${\mfr B}$ -- if ${\mfr A}$ implements the unitary
$U_A$, and ${\mfr B}$ implements $U_B$, then ${\mfr A}\|{\mfr B}$ implements
$U_A\otimes U_B$.  Serial concatenation means the output of ${\mfr A}$
is fed into the input of ${\mfr B}$ -- that is, ${\mfr A}\circ{\mfr B}$ 
implements the unitary $U_BU_A$. 

As an example, consider the pattern 
\begin{equation}\label{pat1}
{\mfr J}_\alpha:=X_2^{s_1}M_1^{-\alpha}E_{12}N_2^0,
\end{equation} 
with $(V, I, O) = (\{1,2\},\{1\},\{2\})$. Given an arbitrary state $\rho$ on qubit $1$, 
this sequence of operations implements $J_\alpha:=HZ(\alpha)$ on the input state and 
places the resulting state $J_{\alpha}\rho J^\dagger_{\alpha}$ on qubit $2$. 
This is one of the fundamental building blocks for 1WQC~\cite{generator04}, 
since it allows for arbitrary one qubit rotations. Any of the local unitaries considered can 
be merged with a (destructive) measurement as follows:
\begin{eqnarray}
M_i^\alpha Z_i(\beta)&=&M_i^{\alpha-\beta}\label{rule1}\\
M_i^\alpha X_i&=&M_i^{-\alpha}\label{rule2}
\end{eqnarray}
and it is readily seen that the $\mfr J_{\alpha}$ pattern above is the serial concatenation of a $Z(\alpha)$ rotation with a modified one-bit teleportation (implementing $H$) -- a well known result for 1WQC~\cite{RB01,ZLC00,Nie05}.
Patterns which lie outside 1WQC model can also be expressed in this extended model, such as
\begin{equation}\label{pat2}
{\mfr X}_{\alpha}:=X_3^{s_2}Z_3^{s_1}M_2^{-(-1)^{s_1}\alpha+{\pi\over 4}}M_1^0E_{23}E_{12}N_2^{\pi\over 4}N_3^0
\end{equation}
with $(V, I, O) = (\{1,2,3\},\{1\},\{3\})$, which implements the unitary 
$HZ(\alpha)H=e^{-i{\alpha\over 2}X}=J_{\alpha}J_0$. It follows from the equations above, that this pattern is equivalent to a $Z(\alpha)$ conjugated by a one-qubit teleportation.
The importance of writing the pattern in this form, using the $N_2^{\pi\over 4}$ preparation, becomes clear when measurements are restricted to the $X$ or $Y$ eigenbasis, as will
be discussed later.
 
Other patterns which play an important role are $\ctR {\mfr Z}$, ${\mfr N}$ and ${\mfr M}$, defined as follows:
$\ctR {\mfr Z}:=E_{12}$, with  $(V, I, O) = (\{1,2\},\{1,2\},\{1,2\})$, 
implements the unitary $\ctR Z$; ${\mfr N}:=N_i^0$ implements
initialization of qubit $i$ into the state $\ket{+}$; and,  
${\mfr M}:=M_i^0$ implements the measurement of qubit $i$ in the $\ket{\pm}$ $X$
eigenbasis. These patterns are crucial in order
to fulfill the DiVincenzo criteria~\cite{DiV00}.

The usual protocol for 1WQC requires computation to be performed in
three steps: individual qubit state preparation, entangling operations 
between qubits, and measurement of individual
qubits with feed-forward of outcomes. In order to follow this protocol
for the generalized model, patterns must be put into a {\em standard
form} where any computation can be performed by a sequence of operations in this
order. Note that these steps do not include the application of single qubit
unitaries, but adaptive measurements can be used to address this
absence, since all quantum computations must end with the measurement
of the qubits in order for information to be extracted. Once a pattern is
in standard form, it is convenient to consider the entangled state that 
is prepared for the computation. Such a state can be described by an {\em 
entanglement graph}, with vertices $V$ and edges $(i,j)$ for every
command $E_{ij}$ in the pattern, where the vertices are labelled
with the initial state in which the qubit is prepared.

The process of turning a given pattern into a pattern in standard form
is called {\em standardization}. The rewrite rules needed for this
procedure are simply \eqref{rule1} and \eqref{rule2}, along with conjugation
relation between unitaries, $E_{12}X_1=X_1Z_2E_{12}$, and
$E_{12}Z_1(\alpha)=Z_1(\alpha)E_{12}$, as well as all the free commutation relations between 
operations on different qubits. Simple rewriting theory arguments~\cite{generator04} 
show that by applying the conjugation relations to move all the local unitaries 
towards the left in the pattern, and then by applying \eqref{rule1} and 
\eqref{rule2}, any pattern 
can be put in standard form.

As mentioned previously, PMM is obtained by setting (i) state preparation angles to
$0$ or  $\tfrac{\pi}{4}$, (ii) measurement angles to
$\tfrac{n\pi}{2}$, and  (iii) local unitaries to $X$ and
$Z(\tfrac{n\pi}{2})$.  Two simple facts follow from this: first, PMM
is closed under standardization and concatenation, as can be readily seen from the
merging and conjugation relations above; second,  PMM contains the
patterns $\ctR {\mfr Z}$, ${\mfr J}_\alpha$, ${\mfr X}_{\beta}$, 
${\mfr N}$ and ${\mfr M}$, where $\alpha={n\pi\over 2}$
and $\beta={n\pi\over2}+{\pi\over4}$, as well as their concatenations.
In particular, ${\mfr X}_{\pi\over4}$ allows for an operation outside the Clifford
group while requiring only Pauli measurements.

\begin{coros}
The PMM is approximately universal in the Solovay-Kitaev sense.
\end{coros}
 
This construction of a universal gate set is equivalent to the construction of 
fault-tolerant universal gate sets via teleportation~\cite{ZLC00,GC99}.


\section{Fault-tolerance}
\subsection{Simulation approach}
In reality, physical implementations of any computational model
are susceptible to noise. In principle, such physical implementation
can be made fault-tolerant by encoding the data and the operations
in a manner such that, even after the overhead of such an encoding is
considered, one can efficiently perform computations of arbitrary
size~\cite{AGP05, AB99,Kit97b,KLZ98,Pre98}. The noise model that is usually considered,
and which we restrict ourselves to in this work, is the model of
independent random failure of each of the operations during computation.
One approach to achieve fault-tolerance in 1WQC is by using fault-tolerance in 
the circuit model as a stepping stone. The construction of fault-tolerant 
circuits is well understood~\cite{ZLC00,GC99}, and it is now well known 
that the implementation of such circuits via 1WQC can lead to 
fault-tolerant quantum computation~\cite{AL05,ND05}.  
This can be most simply understood and demonstrated through the idea of 
composable simulations~\cite{AL05,CLN04}, and the same idea carries through 
to the PMM with minor modifications. The main distinction is
that in the PMM, the change of measurement bases dependent on
measurement outcomes corresponds to a local Clifford correction, as
opposed to a local Pauli correction. Thus the noisy simulations through the PMM
will have an error model which consists of random application of
local Clifford operators. However, because of the linearity
of quantum mechanics and the fact that the Pauli group forms a 
basis for all single qubit operators, the errors are still
correctable as in simulations through the 1WQC model. Thus,
simulating fault-tolerant quantum circuits through the PMM
model is also fault-tolerant.


\subsection{Intrinsic fault-tolerance}
We now turn our attention to the possibility of making any PMM computation
directly fault-tolerant, instead of simulating fault-tolerant quantum circuits 
within 1WQC.

1WQC relies on frequent
measurement to implement a desired state evolution, but none of this
information is used towards fault-tolerance in simulation-based
approaches. The opportunity for improved performance becomes evident
once one considers  the well known link between teleportation and
1WQC~\cite{Nie05,CLN04}, and the fact that FTQC in the circuit model
can achieve very high thresholds via extensive use of teleportation
for simultaneous syndrome extraction and state evolution~\cite{Kni05b}.

\subsubsection{Encoded Computation}
Before we consider how syndrome information is to be extracted, we
must consider encoded computation in the PMM. The basic elements of
the PMM are: preparation of qubits in either $\ket{+}$ or
$\ket{+_{\pi\over 4}}$, pair-wise entanglement via $\ctR Z$, and
measurement in the $X$ or $Y$ eigenbases depending on the outcomes of
previous measurements. Given some quantum code, we can consider these
same elements, but in the subspace corresponding to the code chosen --
that is, preparation of a block of qubits in the encoded states above,
encoded entangling operations, and collective measurements in the
encoded eigenbases $X$ and $Y$. The use of
the 7 qubit self-orthogonal doubly-even CSS codes~\cite{Ste96} 
simplifies the problem considerably if the generators of the encoded
Pauli operators are chosen to be $\enc{Z}=Z^{\otimes7}$ and
$\enc{X}=X^{\otimes7}$. In that case, the encoded entangling operation
$\enc{\ctR{Z}}$ is given by the transversal application of $\ctR Z$ gates
between respective  qubits in two blocks -- in the PMM, it is the parallel
concatenation of the pattern $\ctR{\mfr Z}$. Moreover, measurement in the
encoded $X$ and encoded $Y$ eigenbases are performed by measuring each
of the qubits within the code block in the same basis individually,
followed by classical decoding of the outcomes to determine the
encoded outcome. If we consider concatenated encoding using this
7 qubit code, i.e. 
$\enc{X}^{(j)}=\left(\enc{X}^{(j-1)}\right)^{\otimes 7}$ for the $j$th
level of encoding with $\enc{X}^{(0)}\equiv X$ and similar 
relations for $\enc{Z}^{(j)}$, these transversality properties are 
preserved. 

\begin{figure}
\begin{center}
\includegraphics{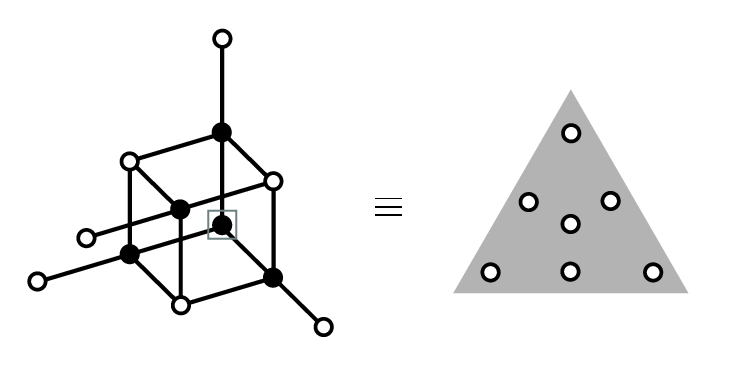}
\end{center}
\caption{Entanglement graph corresponding to the encoding of
a single qubit into the 7 qubit CSS code. The boxed node corresponds to
an arbitrary input qubit. All but the white qubits (corresponding to the
encoding output) are measured in the $X$ basis (up to feed-forward-based
corrections). \label{7-qubit-enc}}
\end{figure}

The encoding procedure of any given stabilizer code over qubits 
is known to correspond to a pattern in 1WQC which allows 
for arbitrary input and requires only measurements along the
eigenbases of the Pauli operators $X$ and $Y$~\cite{SW02,GKR02}
-- this includes both the isomorphism between stabilizer codes and
graph codes, as well as the necessary local Clifford corrections.
If we restrict the inputs to be either $\ket{+}$ or $\ket{+_{\pi\over 4}}$,
we can obtain the encoded states $\ket{\enc{+}}$ or $\ket{\enc{+}_{\pi\over 4}}$
strictly within the PMM. 
The entanglement graph corresponding to the encoding circuit
for the 7 qubit code is depicted in Figure~\ref{7-qubit-enc}. 
Concatenated encoding
proceeds in the obvious way, by serial concatenation of the measurement 
pattern corresponding to the encoding procedure.

However, for the purpose of FTQC, encoding requires verification 
of the encoded states in order to ensure that these state do not 
contain errors that are too correlated~\cite{AGP05,Sho96}.
This can be performed naturally in the PMM via state encoding at
some given level of concatenation, followed by syndrome extracting teleportation
of the lower levels of encoding~\cite{Kni05b}.
There are purification protocols for the entangled
state corresponding to the encoding procedure of any CSS code~\cite{DAB03},
-- such as the 7 qubit code, as depicted in Figure~\ref{7-qubit-enc} --
which may also be employed to reduce errors and error correlations.
We consider only the encoded states that have been successfully verified 
after preparation as part of the computation. 
In this manner, encoded computation in the PMM is 
akin to computation with nested graph states~\cite{Dan05a}, where the 
entanglement graph for encoding is nested within the computation entanglement 
graph. 

It is important to note that the entire concatenated graph state must not be 
purified directly, since the maximum vertex degree of the resulting graph grows linearly
with the level of concatenation, and the purification protocol performance
degrades with higher vertex degrees~\cite{DAB03}. In order to avoid this
problem, one may perform purification per level of concatenation separately,
followed by syndrome extraction teleportation with post-selection
of the states which have a clean syndrome. 

Previous proposals for fault-tolerance in the 1WQC model make use of what is
called the {\it one-buffered implementation} of cluster states~\cite{ND05}.
In such implementations, which are based on the simulation of quantum
circuits, the entanglement subgraph corresponding to the first two time steps in
the circuit model is prepared. The measurements corresponding to the first
time step are performed, followed by the state preparation and entagling operations
corresponding to the third time step of the circuit model. After that, the measurements
for the second time step are performed, and computation proceeds keeping a
one time step ``buffer'' of qubits, so that the entire entanglement graph
need not be prepared in one shot. However, it has been demonstrated
that the 1WQC model, as well as the PMM, allow for greater parallelism
in the computation~\cite{generator04}. In particular, some sequences of operations
which lie in multiple time steps in the circuit model can be performed
in a single time step in these measurement models (a large class of such
operations are unitaries in the Clifford group).
Thus, one may prepare states corresponding to larger subgraphs of
the entanglement graph where all non-output qubits
will be measured simultaneously. There is a partial order constraint for
the timing of the measurements which is implied in the definition of the PMM 
(as well as the 1WQC model), and this partial order gives the dependencies between 
the measurements~\cite{generator04}. One can therefore associate a subgraph
of the entanglement graph with each time step where a collection of measurements
may be performed in parallel. In the case of the PMM, measurement of a
vertex prepared in the $\ket{+_{\pi\over 4}}$ introduces a local
Clifford correction to qubits connected to it in the entanglement graph, and
thus such vertices will always be on the boundary of the subgraphs. However,
patterns implementing Clifford operations have measurements which are
independent of eachother's outcome, and thus the insertion of Clifford
operations in a pattern does not increase the number of such subgraphs, or
equivalently, the minimal number of time steps in which measurements
can be performed in parallel. This is particularly relevant for fault-tolerance, 
as encoding and syndrome extraction operations for stabilizer codes are Clifford operations. 
In principle such operations can be performed in the same time step, if the entire corresponding
subgraph is available for measurement. The preparation of the subgraph itself will require multiple
time step, due to verification, error correction and purification at different
levels of encoding, but since these operations are independent of the rest
of the computation, they may be performed offline.
Clearly, it is not required that maximal parallelism -- corresponding
to the largest subgraph -- be implemented. There is a trade-off between the 
overhead introduced by more complex offline 
preparation and verification of such larger subgraphs, and the lower
effective error rate which may be achieved. Implementations
may range from the one-buffered approach, to the fully parallel approach,
which ensures that all measurements without dependencies can be performed 
simultaneously.

\subsubsection{Syndrome extraction}
In order to perform FTQC, one must be able to
extract information about the errors in the data in order to ensure that 
only clean enough states are introduced into the computation, as described
in the previous section, but also to obtain information about which errors
are likely to have occurred in order to correct them.
This error syndrome extraction can be performed via teleportation, as recently described 
in~\cite{Kni05b,Kni05a}. In essence, the idea is to start with a maximally
entangled pair of encoded qubits $\ket{\enc{\Omega}}_{1,2}=\enc{\ctR{Z}}_{12}\ket{\enc{+}}_1\ket{\enc{+}}_2$ which is prepared offline. 
Given some encoded state $\enc{\rho}$, the error syndrome can be 
extracted in the following manner. Measure each transversal pair 
of physical qubits from $\enc{\rho}$
and the first half of $\ket{\enc{\Omega}}_{1,2}$ in a basis of maximally
entangled states. The state $\enc{\rho}$ is then teleported into the second 
half of 
the entangled pair, up to a tensor product $g$ of local Pauli operators
which is inferred from the outcomes of the pair measurements. 
The error syndrome can in turn be inferred from these corrections by considering
the commutator of $g$ with each of the generators of the stabilizer group of 
the code. This protocol can be seen as the transversal teleportation
of all the physical qubits where the $n$ maximally entangled pairs have
been projected into the codespace being used. Note that this is different
from an encoded teleportation -- an encoded maximally entangled state
is used, but the measurements are performed on physical qubits, not
encoded qubits. This proposed technique for FTQC has not been rigorously 
proven to have an error threshold as is the case for many other 
techniques~\cite{AGP05}, but extensive numerical evidence supports such a
claim~\cite{Kni05b}.

Although the usual teleportation protocol~\cite{BBC93} is performed
with Bell pairs and measurement in the Bell basis, teleportation 
can be performed with any measurement in a basis of maximally entangled 
states, and this choice of 
basis fixes which maximally entangled states can be used as a resource. 
In fact, teleportation
can be performed by the serial concatenation
${\mfr J}_0\circ{\mfr J}_0=X^{s_2}_3Z^{s_1}_3M^0_2M^0_1E_{23}E_{12}N^0_3N^0_2$,
which may be understood as a teleportation using the basis obtained
by applying a Hadamard gate to one of the qubits of a Bell basis. 
If we allow for modified preparation of the entangled resource state,
the pattern, stripped of the entanglement preparation,
simply becomes ${\mfr T}=X^{s_2}_3Z^{s_1}_3M_2^0M_1^0E_{12}$, which, for completeness, must
be concatenated with the pattern for the modified entangled state preparation
(i.e.\ the pattern that prepares the encoded entangled state).

Thus, in the PMM, syndrome extraction of some encoded state $\enc{\rho}$ 
is performed by: (I) preparing and verifying the encoded state 
$\ket{\enc{\Omega}}_{12}$,
(II) teleporting all qubits in $\enc{\rho}$ individually using the
resource state $\ket{\enc{\Omega}}_{12}$, and (III) performing
classical post-processing to infer the syndrome information from
the teleportation measurement outcomes. As discussed, step (I) can
be performed by hierarchical teleportation and 
post-selection~\cite{Kni05b,Kni05a}.
Step (II) can be performed by parallel concatenation of the pattern
${\mfr T}$ above, while step (III) is merely classical post-processing
which affects the bases of subsequent measurements.
Partial syndrome information can be extracted in a similar fashion,
as in the case of the ${\mfr J}_\alpha$ pattern with 
$\alpha={n\pi \over 2}$, where, depending on $\alpha$, one can
obtain information about Pauli errors which anti-commute with $X$
or $Y$. 

\subsubsection{Performing the computation} 
Given any measurement pattern in the PMM, one may make it
fault-tolerant by first translating each of the commands with a larger
pattern representing its encoded form, then inserting instances of the
syndrome extracting teleportation  between each operation, and
standardizing the resulting pattern.

\begin{figure}
\begin{center}
\includegraphics{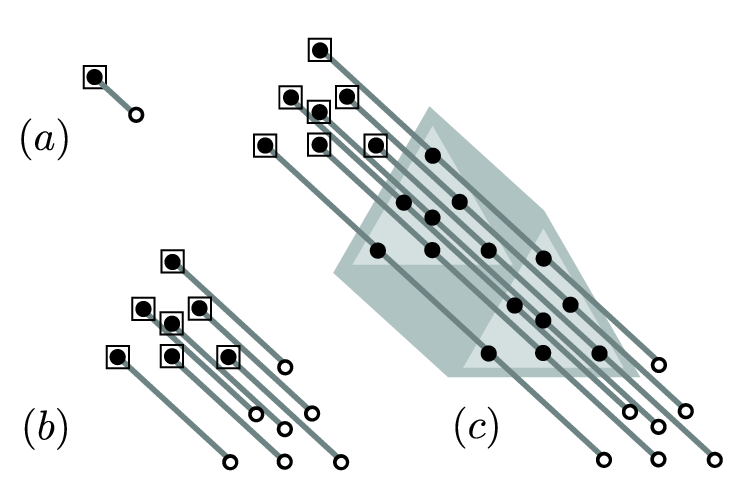}
\end{center}
\caption{Entanglement graphs for the fault-tolerant implementation of
${\mfr J}_0$. The boxed nodes correspond to input qubits, and all
but the white nodes (corresponding to output qubits) are measured
in the $X$ eigenbasis (up to feed-forward-based corrections).\label{h-pat}}
\end{figure} 

As a simple example, consider the pattern  fragment
$X_2^{s_1}M_1^0E_{12}$  that implements the unitary $J_0=H$, with
entanglement graph depicted by Figure~\ref{h-pat}(a). Using a single
level of encoding under the  7 qubit CSS code, the resulting pattern
is already long and omitted for brevity, but its entanglement graph in
Figure~\ref{h-pat}(b) demonstrates  the simplicity of the
transformation. The subgraph enclosed in the shaded triangle
corresponds to the encoded state that must be prepared and verified
before the remaining operations can be performed, in what can be seen
as an extension of the one-buffered implementation of the unencoded
case~\cite{ND05}.  With the data protected by an error correction code
and offline preparation of encoded qubits, one inserts the syndrome
extracting teleportation to obtain the final fault-tolerant pattern
with corresponding entanglement graph depicted in
Figure~\ref{h-pat}(c). Again, the subgraph inside the  irregular
pentagon (corresponding to the preparation of the encoded maximally
entangled pair) is to be prepared and verified before the qubits
within it interact with the remainder of the graph. This demonstrates
the fact that only three subgraphs need to be prepared and verified
offline: the smaller subgraphs corresponding to the encoded states
$\ket{\enc{+}}$ and  $\ket{\enc{+}_{\pi\over 4}}$, and the larger
subgraph corresponding to the encoded state $\ket{\enc{\Omega}}$. This
procedure for implementing fault-tolerance works for any linear graph.
Other graphs, such as the one corresponding to a $\ctR Z$ pattern interacting 
between two linear chains, can be handled in a similar fashion, by simply
inserting syndrome extracting teleportations before and after the
$\ctR Z$ pattern.

It is important to note that the qubits, interactions and
measurements added to the computation in order to extract syndrome
information correspond to Clifford operations on the quantum
states. As pointed out earlier in the paper, the measurements associated 
with a sequence of Clifford 
operations can be performed in any order, even simultaneously and
immediatelly after the qubits are made available for measurement, 
and thus they do not increase the depth complexity of the 
computation~\cite{Rau03,generator04}.
Moreover, this also allows for
the offline preparation of subgraphs corresponding to Clifford
operations, along with measurement of parts of the subgraph, which
allows for the elimination of some types of error via post-selection
-- as pointed out in~\cite{DHN06}, for the case of repeated syndrome
extraction, one can post-select on subgraphs which will yield agreeing
syndromes.

\section{Conclusion}
We have described a measurement based model of computation 
with the notable feature that measurements
are restricted to the eigenbases of the Pauli operators $X$ and $Y$,
and qubit state preparation is extended to both $\ket{+}$ and $\ket{+_{\pi\over 4}}$. 
With the appropriate choice of 
quantum codes, any measurement pattern in this model can be directly
modified into another pattern within the same model, which, 
according to numerical evidence~\cite{Kni05b}, will have a lower 
effective error rate as long as the failure rate per operation
is below a threshold.

After the completion of this work we became aware of similar work 
by Fujii and Yamamoto~\cite{FY06}, where numerical simulations
indicate that the error threshold is comparable with the one
obtained in~\cite{Kni05b}.


\ack
M.S. is partially supported by by NSERC, MITACS and ARO.
E.K. was partially supported by the ARDA, MITACS, ORDCF, and CFI projects during her stay at Institute for Quantum Computing at the University of Waterloo where this work was begun. 

\section*{References}
\bibliographystyle{unrst}

\end{document}